\documentclass[traditabstract]{aa}

\usepackage{dcolumn}
\usepackage{epsf}
\usepackage{txfonts}
\usepackage{graphicx}
\usepackage{booktabs}
\usepackage{rotating}
\usepackage{natbib}
\usepackage{url}
\usepackage{amssymb}
\usepackage{fixltx2e}
\bibpunct{(}{)}{;}{a}{}{,} 

\newcommand{\hi}{\ion{H}{i}}

\def\IRAS{{IRAS\/}}

\def\Herschel{\textit{Herschel}}
\def\Planck{\textit{Planck}}
\def\Wise{\textit{WISE}}

\begin{document}

\title{Probing interstellar turbulence in cirrus with deep optical imaging:\\ no sign of energy dissipation at 0.01 pc scale}

\author{
M.-A. Miville-Desch\^enes\inst{1}, P.-A. Duc\inst{2},  F. Marleau\inst{3}, J.-C. Cuillandre\inst{2,4}, P. Didelon\inst{2},  S. Gwyn\inst{5} and E. Karabal\inst{2,6}
}

  \institute{
 Institut d'Astrophysique Spatiale, CNRS, Univ. Paris-Sud, Universit\'e Paris-Saclay, B\^at. 121, 91405 Orsay Cedex, France 
 \and Laboratoire AIM, Paris-Saclay, CEA/IRFU/SAp - CNRS - Universit\'e Paris Diderot, 91191, Gif-sur-Yvette Cedex, France
 \and  Institute of Astro and Particle Physics, University of Innsbruck, Austria
 \and  Observatoire de Paris, PSL Research University, F-75014 Paris, France
 \and National Research Council of Canada, Herzberg Astronomy and Astrophysics Program, 5071 West Saanich Road, Victoria, BC, V9E 2E7, Canada
 \and  European Southern Observatory,  Karl-Schwarzschild-Str. 2, D-85748 Garching, Germany
 }

 \mail{M.-A. Miville-Desch\^enes: mamd@ias.u-psud.fr}
 \date{\today}

\titlerunning{Probing interstellar turbulence in cirrus with deep optical imaging}
\authorrunning{Miville-Desch\^enes et al.}

\abstract {Diffuse Galactic light has been observed in the optical since the 1930s. We propose that, when observed in the optical with deep imaging surveys, it can be used as a tracer of the turbulent cascade in the diffuse interstellar medium (ISM), down to scales of about 1\,arcsec.
  Here we present a power spectrum analysis of the dust column density of a diffuse cirrus at high Galactic latitude ($l\approx 198^\circ$, $b\approx 32^\circ$) as derived from the combination of a MegaCam $g$-band image, obtained as part of the MATLAS large programme at the CFHT, with \Planck\ radiance and \Wise 12\,$\mu$m data. The combination of these three datasets have allowed us to compute the density power spectrum of the \hi\ over scales of more than three orders of magnitude. We found that the density field is well described by a single power law over scales ranging from 0.01 to 50\,pc. The exponent of the power spectrum, $\gamma=-2.9 \pm 0.1$, is compatible with what is expected for thermally bi-stable and turbulent \hi. We did not find any steepening of the power spectrum at small scales indicating that the typical scale at which turbulent energy is dissipated in this medium is smaller than 0.01\,pc. {The  ambipolar diffusion scenario that is usually proposed as the main dissipative agent, is consistent with our data only if the density of the cloud observed is  higher than the typical values assumed for the cold neutral medium gas. We discuss the new avenue offered by deep optical imaging surveys for the study of the low density ISM structure and turbulence.  }}


\maketitle

\section{Introduction}

In the early 20th century, optical images of star-forming regions revealed nebulae and dark regions indicating the presence of diffuse matter away from stars \citep{kapteyn1909,barnard1910}.
Later, diffuse Galactic light (DGL) was observed in the optical away from the main star-forming complexes, first by \citet{elvey1937} and then by  \citet{de_vaucouleurs1955a,de_vaucouleurs1960,lynds1965}. The DGL is mainly visible towards the Galactic plane but it has been noticed also at high Galactic latitudes \citep{sandage1976}. These early studies rapidly suggested that DGL could be the result of starlight scattered by diffuse matter spread out between stars, that is the interstellar medium (ISM).
Later on, the ubiquity of the ISM was revealed by all-sky surveys at 21\,cm and in the infrared that showed that there is no line of sight on the sky without interstellar matter. 

Following the detection of the infrared emission at high Galactic latitude with \IRAS\ \citep{low1984}, there have been several studies showing the good spatial correlation between the optical DGL and the infrared emission, especially at 100\,$\mu$m \citep{de_vries1985,laureijs1987,guhathakurta1989,paley1991,zagury1999,ienaka2013}.  The strong spatial correlation with the infrared and the spectral energy distribution of the DGL\footnote{The DGL is also observed in the UV \citep[][and references therein]{boissier2015}.} confirmed that it is caused by scattering by large dust grains \citep{brandt2012}.

For extra-galactic studies, one obvious consequence of the presence of dust in the diffuse ISM, referred to as cirrus, is that light emitted in galaxies is partly absorbed and scattered by interstellar dust along the way, even in diffuse areas of the sky.  Moreover, dust does not only produce a chromatic extinction of extra-galactic light, it also scatters photons of the interstellar radiation field towards the observer (the DGL) adding up some brightness to the extra-galactic emission. This is recognized as a nuisance for the study of the diffuse emission in galaxy halos and in between galaxies using deep optical imagery \citep{cortese2010,duc2015}.

Interstellar dust might be a nuisance for extra-galactic studies but, in principle, DGL could also be a useful tracer of interstellar column density, complementing far-infrared and sub-millimeter dust emission \citep{juvela2006}.
In the past decades, to understand the details of the star-formation process, the focus has been on denser and more massive environments like molecular clouds.
For denser regions, the interpretation of scattered light becomes difficult as the observed structure depends on the properties of grains and on the geometry of the radiation field \citep{zagury1999}.
In practice, the UV-visible part of the radiation field gets absorbed efficiently as the column density increases. For studying molecular clouds, near-infrared observation was preferred to optical because of the lower opacity. Nevertheless, the interpretation of near-infrared DGL still requires the use of radiative transfer modeling to trace structure and study the evolution of dust properties \citep{malinen2012,malinen2013}.

Radiative transfer effects are much less of a concern when looking at diffuse regions of the sky, even in the optical.
We notice that optical DGL has not been exploited yet as a way to quantify the structure of the diffuse ISM, even though the {large field of view  cameras  now equipping four-to-eight meter class telescopes offer mapping capabilities with degree size fields of view and with arcsecond angular resolution}. It is difficult to access such a large dynamical range in scales with other observational probes. One potentially interesting application of such high angular resolution observations is to be able to study the properties of interstellar turbulence down to scales close to the dissipation scale.

The formation of dense interstellar structures, where stars form, depends strongly on the way energy and mass is transferred through scales. By creating an energy cascade through more than four orders of magnitude in scale and by producing density fluctuations, interstellar turbulence is a key agent of the overall star-formation process. 
The turbulent cascade of the ISM is more complex than purely hydrodynamical turbulence as it involves compressibility, gravity, thermal instability and magnetic fields which are important elements of the dynamics of the ISM. Polarization observations, in the radio from synchrotron and in the visible-infrared-submm from dust, show a link between the structure of the ISM and the morphology of the magnetic field \citep{planck_collaboration2014a,planck_collaboration2014g,planck_collaboration2014j,planck_collaboration2015b,planck_collaboration2015c}. The \Planck\ observations are compatible with one expected behavior in which matter flows along the field lines, where there is less resistance \citep{soler2013}. In the diffuse ISM, matter is seen to be organized parallel to the field lines while dense and gravitational bound filaments are more perpendicular to the field lines.

The magnetic field is involved in the structure of matter but it is likely that it is also playing a dominant role in the dissipation of the turbulent energy. As the ISM is partly ionized, even in the dense regions, there is friction between the ions and neutrals known as the ambipolar diffusion. This process dissipates energy at a scale larger than the more classical molecular viscosity.
Recently the analysis of Herschel observations have shown that the main filaments of several close-by molecular clouds have a typical width of 0.1\,pc irrespective of their distance and of their column density \citep{arzoumanian2011}. \citet{hennebelle2013b} and \citet{hennebelle2013c} suggested that this might be the signature of the dissipation of turbulent energy by ion-neutral friction.
Alternative explanations have been proposed, for example by \citet{fischera2012} who suggested that 0.1\,pc is a natural scale for a self-gravitating filament in pressure equilibrium with its surrounding.
It is important to note that even if the typical scale found in filaments of molecular clouds might be related to the dissipation of turbulent energy, such a scale has not been identified yet in any spectral analysis of the gas or dust column density \citep[e.g., see][]{miville-deschenes2010}.

Here we propose to use the DGL to study the structure of the diffuse ISM where gravity does not play a dominant role in the dynamics. One advantage of looking at diffuse areas of the sky, away from star-forming regions, is that dust is uniformly heated by the interstellar radiation field, limiting the effects of radiative transfer in the observed structure.
We suggest that large optical telescopes equipped with large field of view camera provide a unique opportunity to study the properties of the interstellar turbulence cascade by observing close-by clouds.
For example, in good seeing conditions, the MegaCam instrument on the CFH telescope can achieve an angular resolution of 0.5\,arcsec. For clouds at a distance of 100\,pc, typical for clouds in the solar neighborhood, this corresponds to a linear physical scale of $2\times 10^{-4}$\,pc, or about 50\,AU. The use of such observations opens a new perspective on the study of the small-scale structure of the ISM and potentially on the identification of the dissipative processes at play.

In this paper we present CFHT/MegaCam observations of a diffuse region at high Galactic latitude. These data are complemented by \Planck\ and \Wise\ data in order to estimate the power spectrum of density fluctuations of the diffuse ISM. The paper is organized as follow. In Sect.~\ref{sec:data} we present the data used in this study. 
Section~\ref{sec:pk_nh} presents the power spectrum analysis. We discuss our results in Sect.~\ref{sec:discussion}.

\begin{figure*}
\begin{center}
\includegraphics[width=16cm, draft=false]{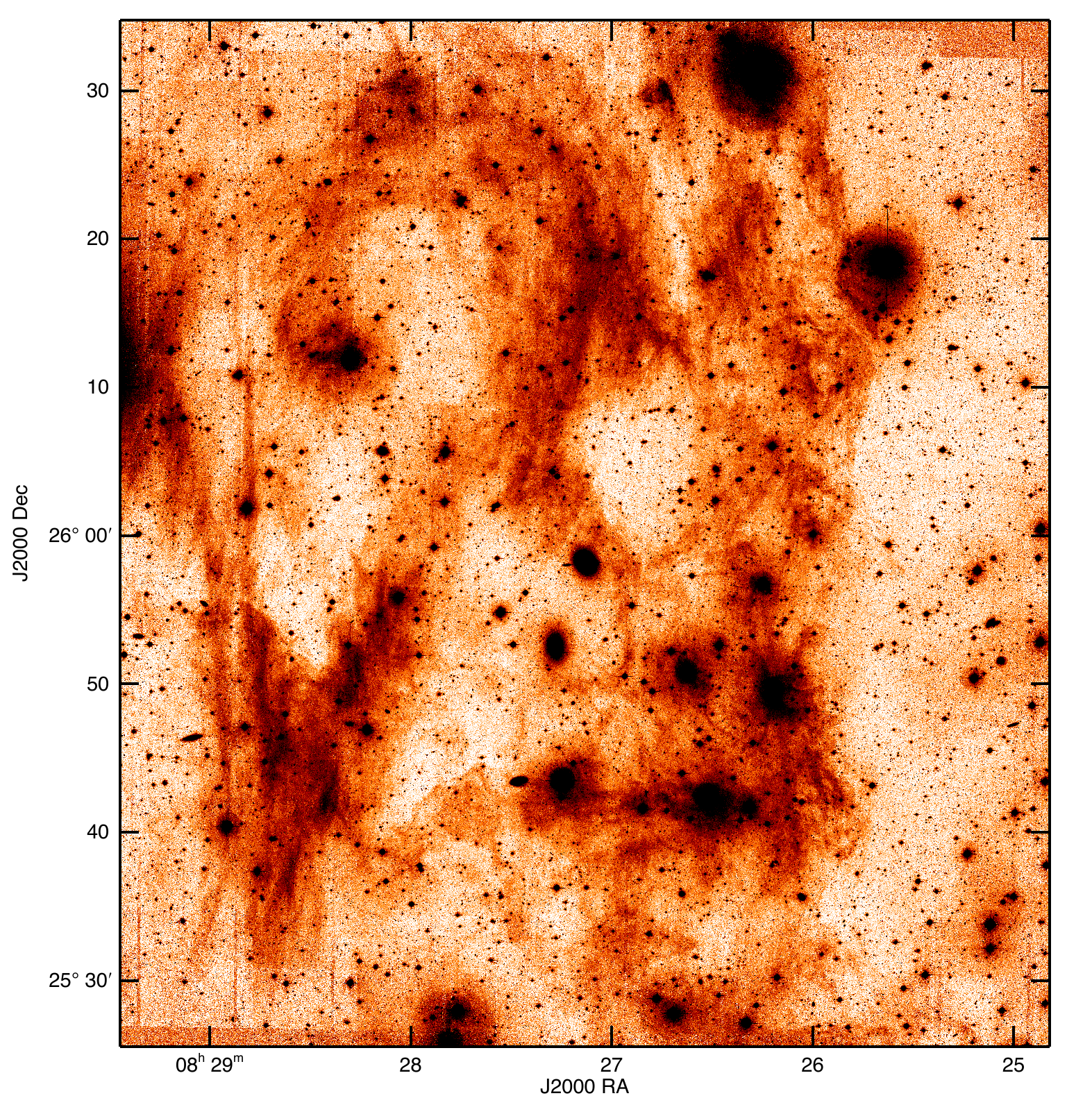}
\includegraphics[width=7.5cm, draft=false]{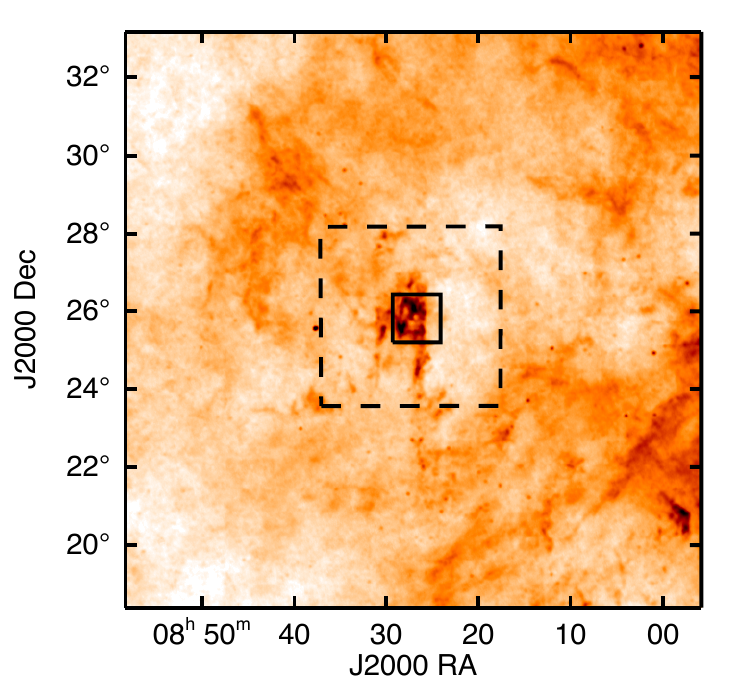}
\includegraphics[width=7.5cm, draft=false]{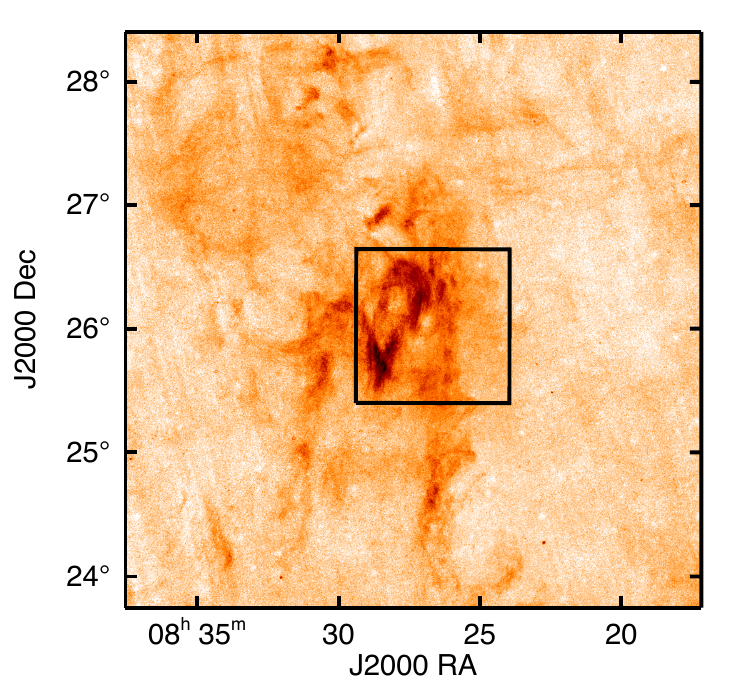}
\caption{\label{fig:maps} Top: MegaCam $g$-band image of the field around NGC\,2592 and NGC\,2594. {The two galaxies are located close to the center of the image}. Bottom-left: \Planck\ radiance. Bottom-right: \Wise\ 12\,$\mu$m. The black rectangle indicates the area of the MegaCam field. The dashed-line rectangle on the radiance map indicates the area of the \Wise\ field shown on the right.}
\end{center}
\end{figure*}

\section{Observing interstellar dust with large optical telescope}

\label{sec:data}

The processing of optical observations of external galaxies is traditionally done in such a way that emission at angular scales larger than the typical size of galaxies is removed, in order to reject spatial variation of the sky background. By doing this, most of the DGL is filtered out.
{Furthermore, images suffer from residual instrumental scattered light. However a substantial gain in the ability of restoring the diffuse emission on large angular scales was obtained using ultra deep and long exposures on amateur-type small telescopes \citep{martinez-delgado2010}  or dedicated observing strategies on large telescopes and processing the data with optimized  pipelines such as Elixir-LSB \citep{ferrarese2012,duc2015}.  These efforts were initially motivated by the detection around nearby galaxies of low surface-brightness collisional debris such as tidal tails. At the surface brightness limit reached by these surveys (typically around 29 mag.arcsec$^{-2}$), images are often polluted by extended patchy diffuse emission from Galactic cirrus. 
Here we exploit one such image, obtained as part of the MATLAS project, a  large program with the MegaCam camera installed at the Canada France Hawaii Telescope, aimed at obtaining deep multi-band images of a complete sample of early-type galaxies from the  ATLAS$^{\rm 3D}$ sample \citep{cappellari2011}. A dithering method with large offsets was used. Final images consist of stacks of seven individual images, with total exposure time of 40 min \citep{duc2015}. 
The target is the field of  the galaxies NGC\,2592 and NGC\,2594, located at $l=197.9^\circ$, $b=32.2^\circ$.
The g-band map of this field is shown in Fig.~\ref{fig:maps}. The angular resolution of this image is 1.3'' and field of view  $1.04^\circ \times 1.15^\circ$.
It shows prominent filamentary but  mostly patchy diffuse emission with structures at all scales. }

\citet{duc2015} already identified the very good match between the diffuse emission of the NGC\,2592-2594 field and the dust emission detected by \Planck-HFI at 857\,GHz (350\,$\mu$m), which leaves little doubt on the DGL nature of the observed filaments. As shown in Fig.~\ref{fig:maps}, the correspondence is also very good with the radiance from \Planck\ \citep{planck_collaboration2014h} and with the \Wise\ 12\,$\mu$m emission \citep{meisner2014}.

The radiance map provided by \citet{planck_collaboration2014h} has an angular resolution of 5'. It is thought to be the best estimate of dust column density in the diffuse areas of the sky, where ${\rm E(B-V)} < 0.3$ (or $A_{\rm V} < 0.93$) and where the interstellar radiation field intensity is uniform, which is the case here. The main advantage of the radiance is that it provides a way to limit the impact of the cosmic infrared background (CIB) anisotropies that degrade strongly the measure of dust optical depth at a given frequency (see \citet{planck_collaboration2014h} for details).  The main structures seen in the MegaCam $g$-band image are also visible in the radiance despite the lower angular resolution.\footnote{The angular resolution of MegaCam is about 300 times better than that of \Planck.} The radiance, shown in Fig.~\ref{fig:maps} for a  $15^\circ\times15^\circ$ region centered on NGC~2592, reveals the extended diffuse emission in the region highlighting the fact that the main structure seen with MegaCAM is slightly brighter than average.

The \Wise\ 12\,$\mu$m map reprocessed by \citet{meisner2014} has an angular resolution of 15'', intermediate between \Planck\ and MegaCam.
A $4.7^\circ\times4.7^\circ$ region centered on NGC~2592 is shown in Fig.~\ref{fig:maps}.
The 12\,$\mu$m emission is dominated by the emission from the smallest dust grains (often attributed to polycyclic aromatic hydrocarbons - PAHs).
It has been shown that the abundance of these small dust grains evolves rapidely in the ISM through fragmentation and coagulation processes \citep{miville-deschenes2002}. In molecular clouds, the emission from these small dust grains tends to be lower, probably because they stick to bigger grains \citep{stepnik2003}.
Nevertheless, in a uniformly illuminated region of relatively small volume density, like the high Galactic cloud studied here, it is likely that the PAH emission is proportional to the gas column density, at least to first order. In that respect, the excellent spatial correspondance of the \Wise\ map with the radiance from \Planck\ seen in Fig.~\ref{fig:maps} is reassuring. It is important to mention that in their reprocessing of the \Wise\ 12\,$\mu$m map, \citet{meisner2014} used the \Planck\ 857\,GHz data to recover the emission at scales larger than $2^\circ$.

\begin{figure}
\begin{center}
\includegraphics[width=0.88\linewidth, draft=false, angle=0]{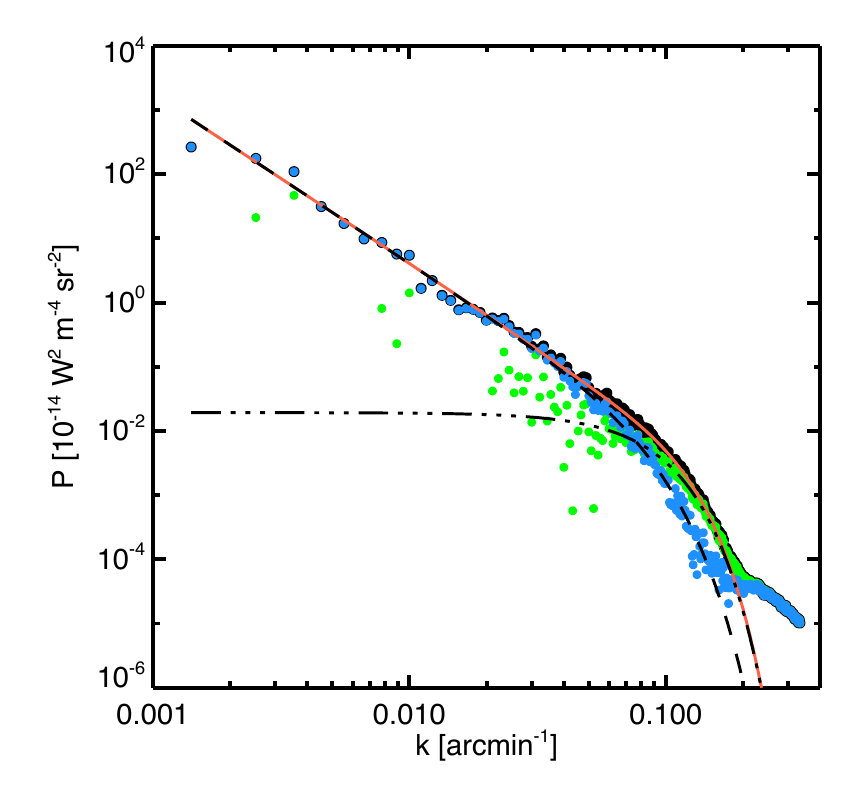}
\includegraphics[width=0.88\linewidth, draft=false, angle=0]{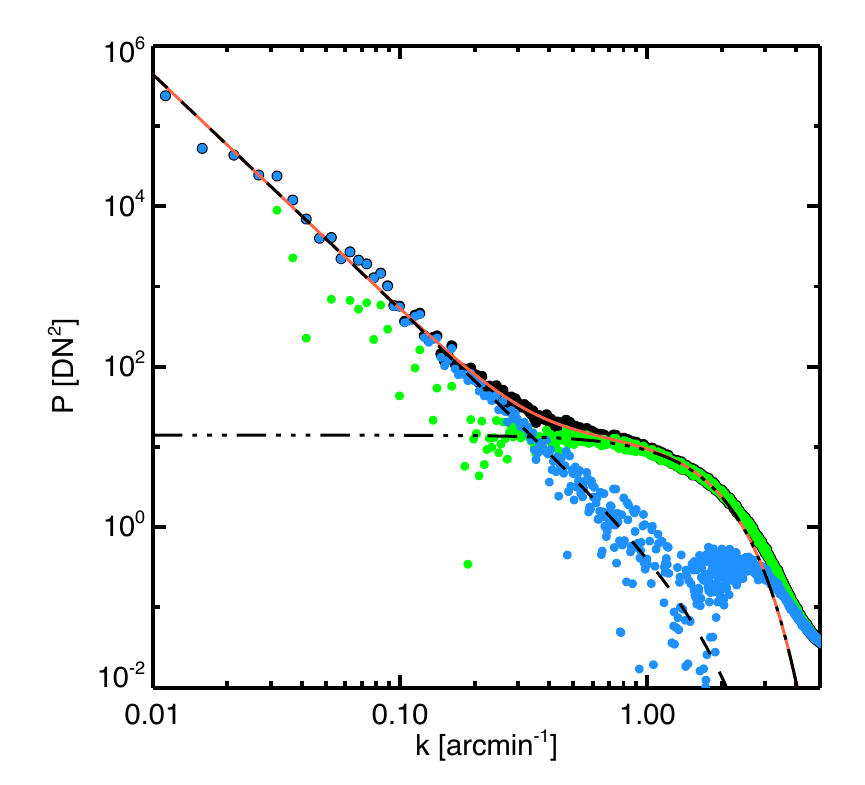}
\includegraphics[width=0.88\linewidth, draft=false, angle=0]{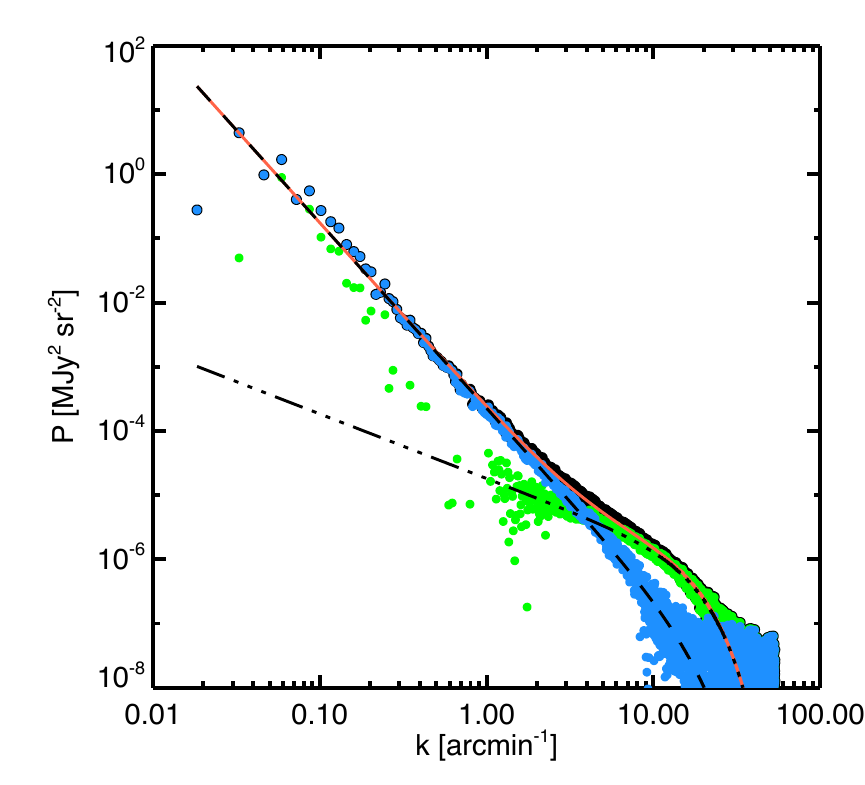}
\caption{\label{fig:pk_indiv} Power spectrum model of the individual maps : \Planck\ (top), \Wise\ (middle) and MegaCam (bottom). The power spectrum is given by the black dots. The global fit is the red curve.  The blue points represent the cirrus component ($P(k) - B(k) \times C k^\beta$) while the green dots represent the noise component ($P(k) - B(k) \times A k^\gamma$). The dashed and dash-dotted curves are the results of the fit for the cirrus and noise terms respectively. The slope of the dust emission power spectrum is $\gamma=-2.7\pm0.1$,  $\gamma=-2.9\pm0.1$ and  $\gamma=-2.9\pm0.1$ for \Planck, \Wise\ and MegaCam, respectively.}
\end{center}
\end{figure}

\section{Power spectrum of column density}

\label{sec:pk_nh}

\subsection{Context}

One way of exploring the turbulent cascade and the associated dissipative processes is to study the statistical properties of column density fluctuations of interstellar clouds. In particular the power spectrum of column density is bearing information about the way density is organized in three dimensions and therefore about the very nature of the turbulent flow. Several studies showed that the power spectrum of the column density of an optically thin tracer is equal to the power spectrum of the density \citep[e.g.,][]{miville-deschenes2003b}.
If the power spectrum of the column density provides some information on the inertial range of the turbulence and on the type of the turbulent flow, it is plausible that the dissipation of the turbulent energy will also leave an imprint. Typically, the dissipation of the turbulent energy induces a loss of power in the velocity field at scales of the order of the dissipation scale. Because density and velocity fluctuations are related, it is expected that the power spectrum of the density field (and therefore of the column density field) will show a similar loss of power at small scales. This effect is clearly seen by \citet{ntormousi2016} in their analysis of numerical simulations dedicated to the study of the imprint of ambipolar diffusion on the statistical properties of the diffuse ISM.

\begin{figure}
\begin{center}
\includegraphics[width=\linewidth, draft=false, angle=0]{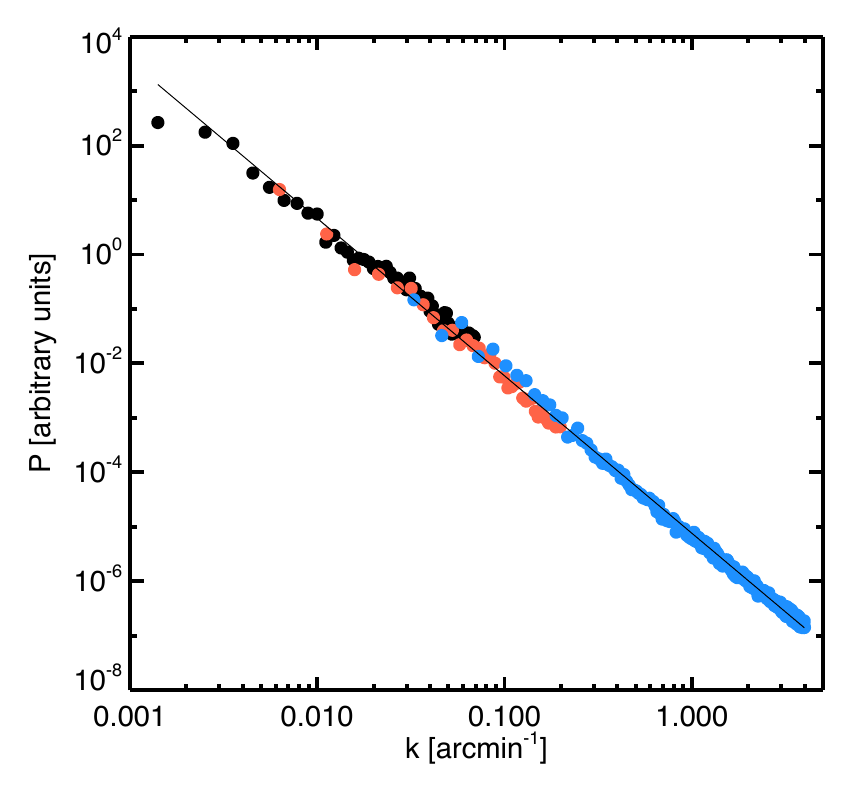}
\caption{\label{fig:pk_log} Combined power spectrum showing the three $P(k)$: black is \Planck\ radiance, red is \Wise\ and blue is MegaCam. The units of the y axis are arbitrary; each $P(k)$ was scaled in order to match the others. For each $P(k)$, we show data points corresponding to scales larger than the beam and where the power is above the noise component. The data points shown here are noise subtracted and divided by the beam function. The best fit gives $P(k) \propto k^{-2.9\pm0.1}$.}
\end{center}
\end{figure}

Based on this idea, the present study focuses on the characterization of the power spectrum of column density fluctuations of the NGC\,2592/2594 field.
In principle, by combining the MegaCam data with \Planck\ and \Wise\ data, we are able to study the statistical properties of the density fluctuations on scales ranging from $\sim 1$'' to $15^\circ$.
Obviously, the combination of these three data sets can not be done everywhere on the sky as they trace dust in different ways: the big grain emission (\Planck), the PAH emission (\Wise) and the scattering from big grains (MegaCAM). In the very diffuse area of the sky studied here, where there is no local heating source and where the column density is in a range where radiative transfer and dust evolution effects are minimal, these three observations are assumed to be reliable tracers of the dust and gas column density. 

\subsection{Star and galaxy removal}

One obvious aspect of the $g$-band data is the presence of a large number of point-like sources, that is, stars and distant galaxies, and extended nearby galaxies, including NGC\,2592/2594. These sources dominate the power in the map. They must be removed in order to study the power spectrum of the Galactic diffuse emission. To do so, we used initial  catalogs of individual sources computed by  
  {\em SExtractor}  \citep{bertin1996}, using different values of the  {\em isoarea\_image} parameter.
The affected pixels were then  replaced by a median value of the surrounding background. In the process, 8.7\% of pixels were modified.

Figure~\ref{fig:maps} shows the quality of the MegaCam data processing.  In this image, the identified individual sources that were masked  in the final image appear as black. The  footprint of the individual MegaCam images is barely visible.  The vertical lines at the  top and bottom of the image are part of a residual  instrumental signature. Apart from them, all features visible in the image are real, including the multiple very narrow filaments.
The small-scale structure of the DGL is thus revealed with striking details.

\subsection{Power spectrum analysis}

\label{sec:pk}

The power spectrum of the three maps (MegaCam, \Wise\ and \Planck\ radiance) was computed using a standard technique \citep[see][]{miville-deschenes2002b,miville-deschenes2010}. 
The power spectrum, $P(k)$, is the azimuthal average of the modulus of the Fourier transform of each image.
To avoid edge effects in the $X$ and $Y$ direction due to the replication of the image done by the fast Fourier transform algorithm, the edges of the images were first slightly apodized using a cosine function. This has almost no effect on the result as the three maps show very small large scale gradients. 

The power spectrum of each map was modeled using the expression
\begin{equation}
P(k) = B(k) \times \left[ A k^\gamma + C k^\beta \right].
\end{equation}
Here we assumed that the power spectrum of each map is well represented by two components convolved by a Gaussian beam, $B(k)$. The first contribution to the power spectrum, $A k^\gamma$, is the interstellar dust that is assumed to follow a power law. The second contribution, $C k^\beta$, is included to represent the contribution from point sources and from the noise. In this type of analysis, this second component is often assumed to be white ($\beta=0$) but it can also be gray ($\beta < 0$) depending on the statistical properties of the noise, potential instrumental effects or residual contamination emission (1/f noise has $\beta=-1$). The FWHM of the Gaussian beam was set to 5', 15'' and 1.3'' for \Planck, \Wise\ and MegaCam respectively.

The power spectrum of each map is shown in Fig.~\ref{fig:pk_indiv} (top \Planck\ radiance, middle \Wise, bottom MegaCam). For each panel, the black points represent the power spectrum of each map. The global fit is the red curve. The cirrus (blue points and dashed curve) and noise components (green points and dash-dot-dot-dot curve) are also shown.-

The three maps have a power spectrum that is well described by a power law at large angular scales (low values of $k$). The slope of the cirrus component is almost the same in the three maps: $\gamma=-2.7\pm0.1$,  $\gamma=-2.9\pm0.1$ and  $\gamma=-2.9\pm0.1$ for \Planck, \Wise\ and MegaCam respectively.

For \Planck\ and \Wise, the noise term is well fit with $\beta=0$, compatible with white noise or with CIB anisotropies. This term is extremely small for \Planck\ in accordance with the fact that CIB anisotropies are small in the radiance map. For \Wise\ the noise term is significant; it reveals the limited signal-to-noise ratio of the data in such diffuse areas of the sky.\footnote{\citet{meisner2014} convolved the data and thus the noise to 15'' resolution.}
Contrary to the two other maps, the noise term of MegaCam is not white. The best fit is compatible with $\beta=-1.01$ (1/f noise). The structure in the noise is likely dominated by residual emission from individual sources removal.

The similarities in the cirrus power spectrum for the three maps are better seen in Fig.~\ref{fig:pk_log} where they are combined in the same plot. Here the units are arbitrary; each $P(k)$ was scaled in order to match the others on the common scales. For each $P(k)$, we show only the data points corresponding to scales larger than the beam and where the power is above the noise component.

\section{Discussion}

\label{sec:discussion}

\subsection{Where is the cirrus optical emission coming from?}

\begin{figure}
\begin{center}
\includegraphics[width=\linewidth, draft=false, angle=0]{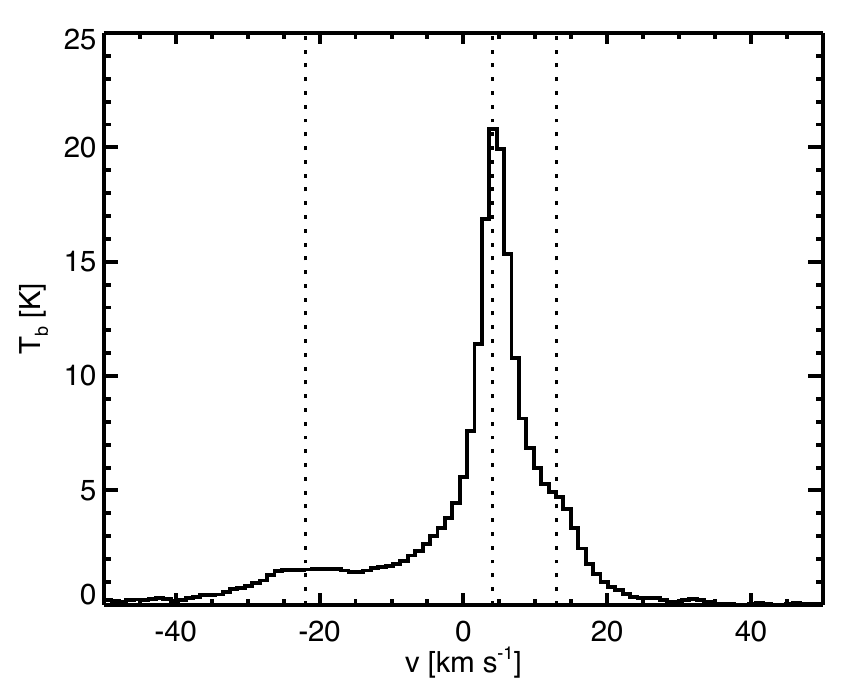}
\caption{\label{fig:21cm_spectrum} 21\,cm spectrum at the central position of the NGC\,2592 ATLAS field. Data are from the LAB survey \citep{kalberla2005}. The dashed lines ($v_{\rm LSR}= -21.6$, $4.1$ and $13.4$\,km\,s$^{-1}$) correspond to the channel maps shown in Fig.~\ref{fig:21cm_channel}.}
\end{center}
\end{figure}

Before discussing the results of the power spectrum analysis, it important to estimate the distance to the extended diffuse structure observed with MegaCam. To our knowledge, there is no direct measure of distance to the specific cirrus structure studied here. 
Estimating the exact distance to the cloud is difficult without absorption measurements of stars located behind it. According to the 3D map of the dense clouds around the Sun produced by \citet{lallement2014}, there is very little dense material in the direction of NGC~2592-2594 ($l=197.9^\circ$, $b=32.2^\circ$). It is therefore difficult to accurately estimate the distance to the cloud.

\begin{figure}
\begin{center}
\includegraphics[draft=false, angle=0]{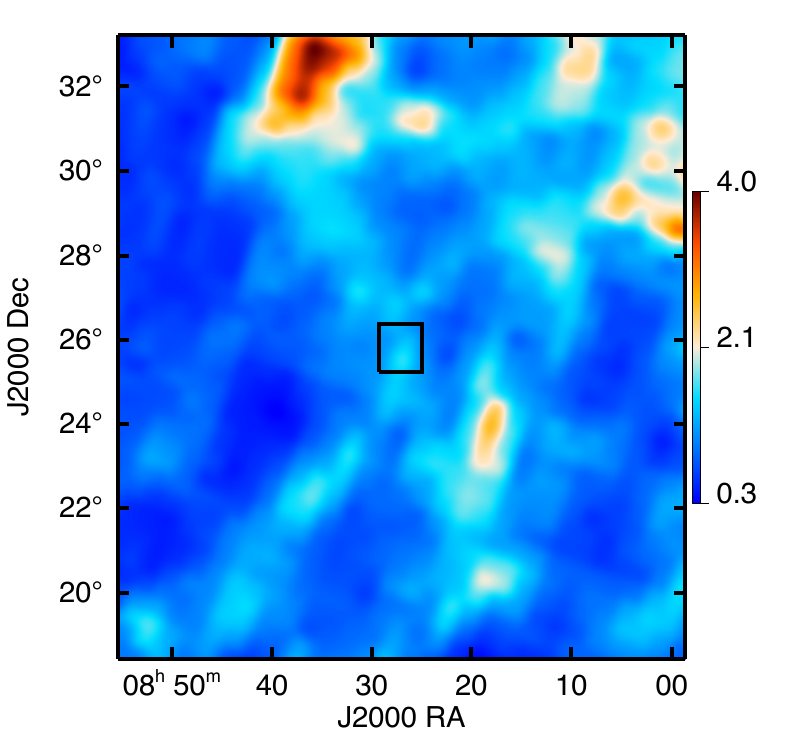}
\includegraphics[draft=false, angle=0]{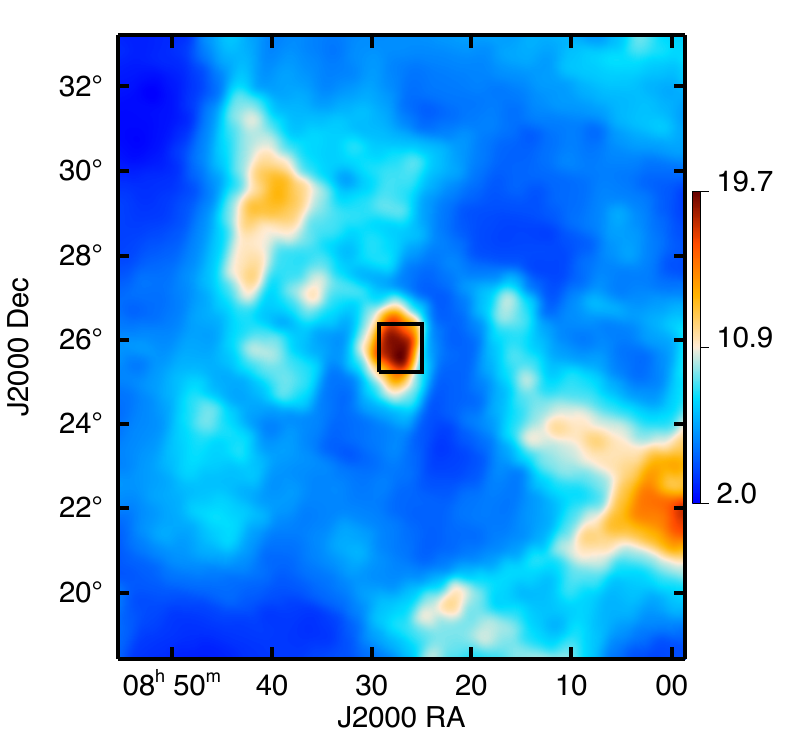}
\includegraphics[draft=false, angle=0]{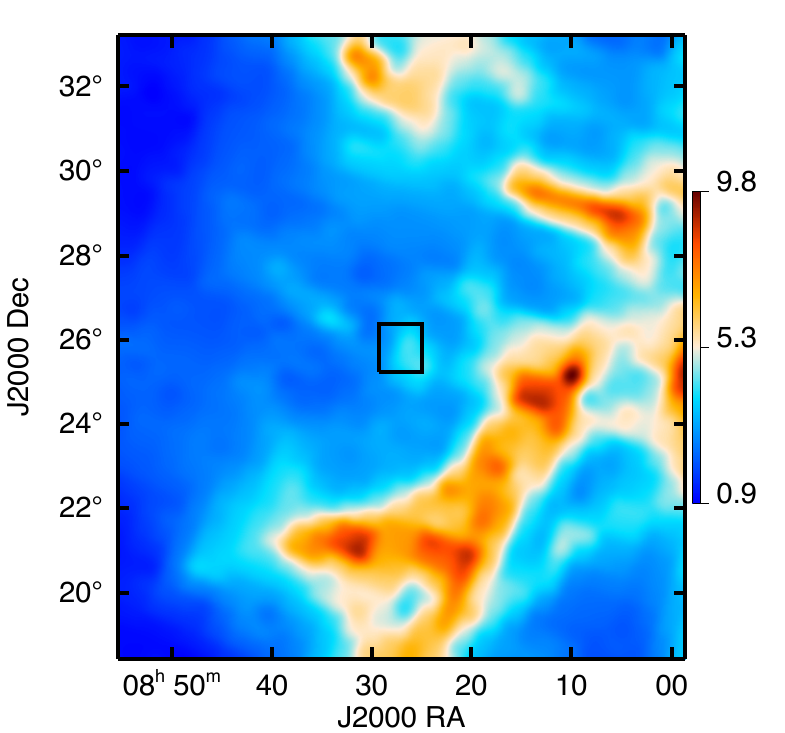}
\caption{\label{fig:21cm_channel} Channel maps of 21\,cm brightness temperature for the NGC\,2592 area. Data are from the LAB survey \citep{kalberla2005}. Units are $K$. The three channels maps are for $v_{\rm LSR}= -21.6$\,km\,s$^{-1}$ (top), $4.1$\,km\,s$^{-1}$ (middle) and $13.4$\,km\,s$^{-1}$ (bottom). They correspond to the dashed lines of Fig.~\ref{fig:21cm_spectrum}. We note the different dynamic range of each map. The area shown is the same as for the \Planck\ radiance (Fig.~\ref{fig:maps}). The black square indicates the MegaCam field.}
\end{center}
\end{figure}

Given that the field is at high Galactic latitude ($b=31.6^\circ$), one possibility could be that the structure observed is part of the Galactic halo. In that case it is probable that it would fall into the Intermediate-Velocity range. To evaluate this we looked at 21\,cm from the Leiden/Argentine/Bonn (LAB) survey \citep{kalberla2005}. Even with its coarse angular resolution of 30', the average velocity of the 21\,cm emission  provides a mean to identify if the emission is coming from the local ISM or if it is located in the Galactic halo. Figure~\ref{fig:21cm_spectrum} shows the 21\,cm emission spectrum at the central position of the MegaCam field. It reaches $T_b \approx 21$\,K and the \hi\ column density is $N_{\rm HI} = 4.3\times10^{20}$\,cm$^{-2}$, both estimated within a 30' beam. The main structure seen in the dust tracers (MegaCam, \Wise\ and \Planck) is seen at a 21\,cm velocity of about 4\,km\,s$^{-1}$ (Fig.~\ref{fig:21cm_channel} - middle). Therefore it is unlikely to be located in the Galactic halo where clouds usually have negative velocities as they slow down when entering the disk. 

The high Galactic latitude and the fact that it is seen at 21\,cm as a narrow cold neutral medium (CNM) component at a LSR velocity close to zero, are indications that the observed structure is in the solar neighborhood. It is a fairly compact structure, less than one degree across. It could be a small, dense cloud that was not identified in the 3D map of \citet{lallement2014} based on a limited number of sightlines to background stars. 
We will make the assumption that the cloud is likely to be located within the thin \hi\ disk of HWHM=106\,pc \citep{dickey1990}. At the latitude of the cloud, this translates into a probable distance of 200\,pc, a value we adopt here.

\begin{figure}
\begin{center}
\includegraphics[width=\linewidth, draft=false, angle=0]{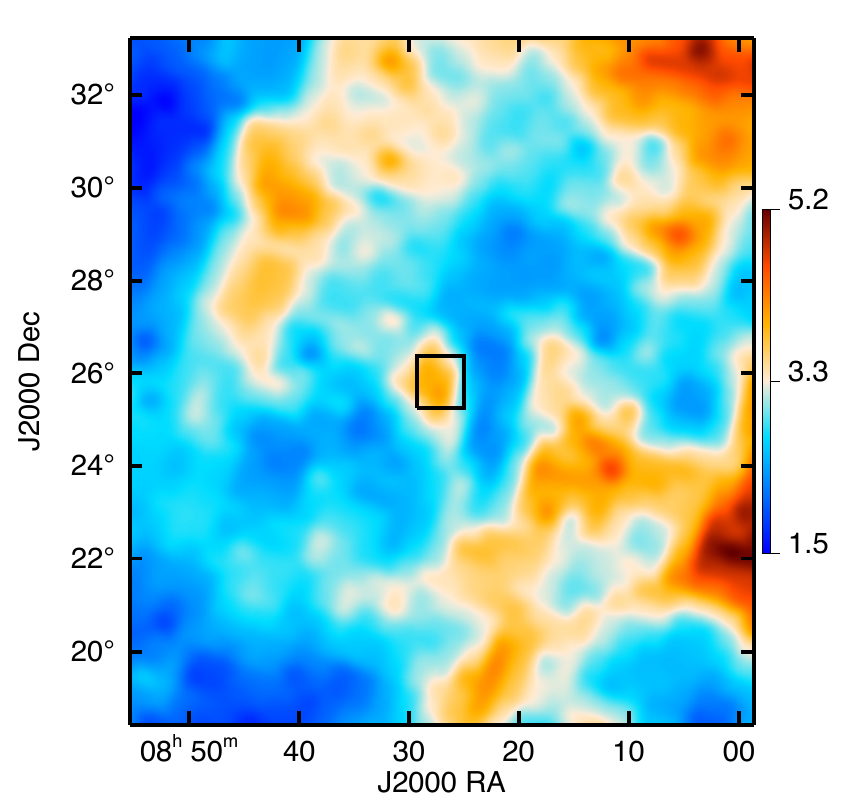}
\caption{\label{fig:21cm_NH} Map of the \hi\ column density estimated from the 21\,cm data of the LAB survey \citep{kalberla2005}. The units are $10^{20}$\,cm$^{-2}$. The area shown is the same as for the \Planck\ radiance (Fig.~\ref{fig:maps}). The black square indicates the ATLAS field.}
\end{center}
\end{figure}

\subsection{The power spectrum of density fluctuations}

The combined power spectrum presented Fig.~\ref{fig:pk_log} indicates that the power spectrum of density fluctuations in cirrus follows a power law with a constant exponent, $\gamma=-2.9 \pm 0.1$, over three orders of magnitude in scales. Our analysis does not reveal any characteristic scale, neither at large nor small scales. 
The exponent found here is comparable to what was found earlier using 100\,$\mu$m \IRAS\ data, $\gamma \sim -2.9$, \citep{gautier1992,miville-deschenes2007a} or 250\,$\mu$m \Herschel-SPIRE data on a similar range of scales (down to 30''), $\gamma=-2.7\pm0.1$, \citep{miville-deschenes2010}. Our results are also similar to what was found recently using 21\,cm observations of diffuse high Galactic latitude fields (values ranging from $-2.5$ to $-3.0$) by \citet{martin2015} and \citet{blagrave2016}.

Even though no characteristic scale is found in the power spectrum, the exact value of $\gamma$ provides some information about the properties of interstellar turbulence.
For compressible, isothermal and subsonic turbulence, \citet{kim2005,kritsuk2007,saury2014} showed that the power spectrum of density follows that of the velocity ($P_k(n) \sim k^{-11/3}$). This would produce a column density power spectrum with $\gamma = -3.7$, far from what is observed. For compressible turbulence, the increase of the Mach number into the supersonic regime leads to larger small-scale density fluctuations and therefore to a flatter power spectrum of the (column) density. According to \citet{kim2005}, a density power spectrum with $\gamma=-2.9$ would correspond to an isothermal flow with a Mach number $M\sim 7$.

Similarly a flatter power spectrum is also observed in the case of thermally bi-stable gas \citep{gazol2010} even when the turbulence is subsonic \citep{saury2014}.
In particular, \citet{saury2014} showed that the thermal instability generates high density contrasts, without having to rely on the effect of supersonic turbulence. In their simulation they showed that for typical properties of the WNM in the solar neighborhood, the density field of the \hi\ (CNM plus WNM) has a power law power spectum with $\gamma \sim -2.5$. 

The 21\,cm data in this area of the sky is typical of high Galactic latitude cirrus. The 21\,cm line emission (Fig.~\ref{fig:21cm_spectrum}) is characterized by a narrow component (FWHM$\sim 5$\,km\,s$^{-1}$) on top of a wide one (FWHM$\sim 20$\,km\,s$^{-1}$), a classical picture of the thermally bi-stable \hi. This is compatible with the fact that the density structure is the result of thermal instability of the \hi.

\begin{table}
  \caption{\label{tab:ldiss} Typical dissipation scales of ambipolar diffusion for different interstellar conditions.}
  \tabskip=0pt
\begin{center}
\begin{tabular}{ccccc}\specialrule{\lightrulewidth}{0pt}{0pt} \specialrule{\lightrulewidth}{1.5pt}{\belowrulesep}
  Medium & $n_n$ & $B$ & $X$ & $l_{\rm AD}$ \\
  & [cm$^{-3}$] & [$\mu G$] & & [pc] \\ \midrule
  molecular clouds & $10^4$ & 80 & $1\times10^{-7}$ & 0.1\\
  cold \hi & 40 & 6 & $1\times 10^{-4}$ & 0.02\\
  warm \hi & 0.5 & 6 & $3\times 10^{-3}$ & 0.5\\ \bottomrule[\lightrulewidth]
\end{tabular}
\end{center}
\end{table}

\subsection{The ion-neutral friction scale}

The structure observed with MegaCam is likely to be a mixture of CNM and WNM located in the solar neighborhood.
The power spectrum analysis shown in Sect.~\ref{sec:pk} reveals that the cirrus density structure is characterized by a single power law over the whole range of scales. Assuming that turbulence is involved in shaping the multi-scale structure of the density field, it implies that our analysis did not reach the energy injection scale nor the dissipation scale. 
In this context, it is interesting to evaluate what is the smallest physical scale probed by the MegaCam data. Given the presence of a noise component at small angular scale (Fig.~\ref{fig:pk_indiv}-bottom), the scale at which the cirrus and noise components are equal is $k \sim 4$\,arcmin$^{-1}$, corresponding to 15''. Assuming a distance to the cloud of 200\,pc, the smallest accessible scale is 0.01\,pc. 

In the ISM, two processes are considered for dissipating the turbulent energy: molecular dissipation and ambipolar diffusion.
Given the low density of the ISM and the fact that is never totally neutral, it is likely that the dissipation of the turbulent energy is dominated by ion-neutral friction (ambipolar diffusion). 
Following \citet{lequeux2005,miville-deschenes2016} the scale at which ambipolar diffusion becomes significant for a partially ionized medium is given by
\begin{equation}
  \label{eq:ldiss}
l_{\rm AD} = \sqrt{\frac{\pi}{\mu_n}} \frac{B}{2X\,\langle\sigma v\rangle \, n_n^{3/2}}
  ,\end{equation}
where $X$ is the ionization fraction, $n_n$ the density of neutrals, $\langle \sigma v \rangle$ the collision rate between ions and neutral (assumed to be the Langevin rate $2 \times 10^{-9}$\,cm$^3$\,s$^{-1}$) and $\mu_n$ the molecular weight of the neutrals, taken to be $1.4\,m_p$ where $m_p$ is the proton mass.

\begin{figure}
\begin{center}
\includegraphics[draft=false, angle=0]{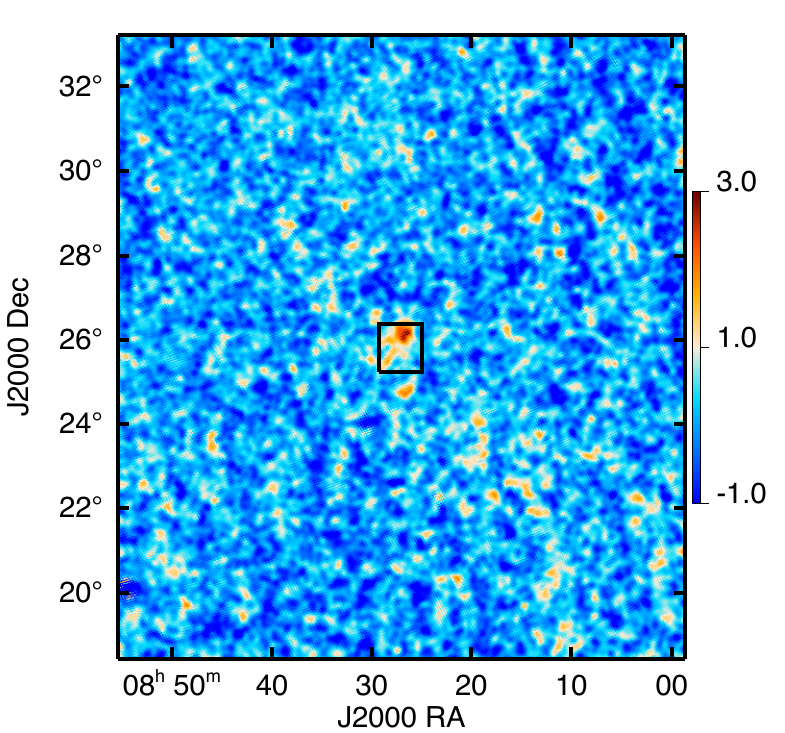}
\caption{\label{fig:CO} Map of CO emission ($J=2-1$) as observed by \Planck. The map covers the same area as the \Planck\ radiance (Fig.~\ref{fig:maps}) and the 21\,cm data (Fig.~\ref{fig:21cm_NH}). There is no detected emission in the $J=1-0$ \Planck\ map. Units are $K_{\rm CMB}$. The black square indicates the ATLAS field.}
\end{center}
\end{figure}

The difficulty in estimating $l_{AD}$ is in defining typical values for $n_n$, $B$ and $X$ as a function of environment or phases. As summarized by \citet{heiles2005}, the total field strength in CNM clouds in the solar neighborhood is, on average, $B_{\rm tot} = 6.0 \pm 1.8$\,$\mu$G. This is also the average field strength value measured in the diffuse warm components of the ISM (WIM, WNM) using synchrotron data \citep{beck2001}. It is therefore assumed that the magnetic field strength does not depend significantly on gas density for $n<100$\,cm$^{-3}$. At higher densities, in molecular clouds, the field strength is observed to scale roughly with the square root of density with $B \approx B_0 \sqrt(n/n_0)$ with $B_0$=25\,$\mu$G and $n_0=10^3$\,cm$^{-3}$ \citep{crutcher1999}.

In the cold \hi, the ionization fraction (or equivalently the number of free electrons) is given by the abundance of ionized carbon (CII) which is the main source of electrons. It is close to $X=1\times 10^{-4}$ \citep{wolfire2003}. In the warm \hi\ the ionization fraction increases to $3\times 10^{-3}$. In molecular clouds $X$ depends strongly on local conditions; it is assumed to be in the range $X=10^{-6}-10^{-7}$ \citep{bergin2007}. 

The density of the molecular gas varies greatly, depending on scale and mass of the clouds, with values ranging from $10^2$ to $10^6$\,cm$^{-3}$ \citep{ferriere2001}. Here we assume $n=10^4$\,cm$^{-3}$. The \hi\ average density is better constrained as it is mostly set by the pressure equilibrium of the two-phase medium \citep{wolfire2003,saury2014}. Here we assume $n(\rm{CNM})=40$\,cm$^{-3}$ and $n(\rm{WNM})=0.5$\,cm$^{-3}$.

Given these physical conditions for molecular clouds, CNM and WNM (summarized in Table~\ref{tab:ldiss}), and using Eq.~\ref{eq:ldiss}, the ambipolar diffusion scales for the three media are 0.1\,pc, 0.02\,pc and 0.5\,pc, respectively. The ion-friction scale is the smallest in the cold \hi\ but the difference with the molecular phase is not immense considering the very different physical conditions. This is explained by the specific dependance of $l_{\rm AD}$ on $n_n$, $B$ and $X$ (see Eq.~\ref{eq:ldiss}); the lower density of the CNM and lower magnetic field strength are almost compensated by the higher ionization fraction. Interstingly, it is in the warmer phase, where the ionization fraction is the highest, that the ion-neutral friction scale is the largest. 

The relatively narrow 21\,cm spectral feature (Fig.~\ref{fig:21cm_spectrum}) associated with the cloud seen in the $g$-band image indicates that a significant fraction of the gas is in the CNM phase. 
We estimate that for density and magnetic field intensity typical of the CNM, the ion-neutral friction scale is about 0.02\,pc. At a distance of 200\,pc this corresponds to 20\,arcsec (or equivalently to 2.9\,arcmin$^{-1}$). This corresponds to an angular scale where the signal-to-noise ratio of the MegaCam data is slightly above one (Fig.~\ref{fig:pk_indiv}) but no steepening of the power spectrum is observed.
We conclude that the CNM dissipation scale is smaller than $0.01-0.02$\,pc.

One explanation might be that the gas observed here is denser than typical CNM gas.
In the bright parts of the MegaCam field, the column density estimated from the radiance \citep[using the conversion factor given by][]{planck_collaboration2014h} is $N_{H} = 3.1\times 10^{21}$\,cm$^{-2}$. On the other hand the column density derived from 21\,cm data peaks at $N_{HI} \approx 4.6\times 10^{20}$\,cm$^{-2}$. Part of this difference is due to the different angular resolution (5' and 30') but it can not explain it entirely. The difference between dust-derived and \hi\ column density is an indication that a significant fraction of the gas is molecular, implying a higher density than usual CNM. There is indeed detectable $^{12}$CO ($J=2-1$) emission in the \Planck\ data at the position of the MegaCam field (Fig.~\ref{fig:CO}). If the gas is denser ($n\sim10^3-10^4$\,cm$^{-3}$) and the amplitude of the magnetic field typical of the diffuse ISM ($B\sim 6$\,$\mu$G), it could explain why the dissipation scale is not visible in the power spectrum of the column density. Another potential explanation is that the cloud is further away than 200\,pc.

\section{Conclusion}

We have presented a power spectrum analysis of the dust column density of a diffuse area at high Galactic latitude ($l\approx 198^\circ$, $b\approx 32^\circ$) as derived from three different tracers: \Planck\ radiance, \Wise 12$\mu$m and MegaCam $g$ band. The combination of these three datasets allowed us to compute the density power spectrum of the CNM over more than three orders of magnitudes in scales. We found that the density field is well described as a single power law over scales ranging from 0.01 to 50\,pc. The power law exponent, $\gamma=-2.9 \pm 0.1$, is compatible with what is expected for the density field of thermally bi-stable and turbulent \hi.

No significant flattening of the spectrum is observed at large scales, indicating that the energy injection scale is likely to be larger than 50\,pc. In addition, we did not find any steepening of the power spectrum at small scales indicating that the typical scale at which turbulent energy is dissipated in this medium is at scales smaller than 0.01\,pc. The comparison of the dust tracers with 21\,cm and CO data is indicative of a CNM density slightly larger than average. In this case, the ambipolar diffusion scale is likely to be smaller than 0.01\,pc.

{ We suggest that optical observations of diffuse high Galactic latitude regions can be used as a tracer of the properties of interstellar turbulence at small angular scale. In good seeing conditions, observations with MegaCAM on the CFH telescope can provide images at a resolution up to 0.5\,arcsec. For clouds at a typical distance of 100\,pc, this angular resolution corresponds to a linear physical scale of $2\times 10^{-4}$\,pc, or about 50\,AU.  The field used in this analysis has a rather  poor seeing but was selected because of its strong and large scale cirrus distribution  that allowed us  to test the feasibility of the method. The MATLAS large program  already provides a high number of images  contaminated by the DGL. It is present in about 20\% of the 241  fields observed at all Galactic latitudes with seeing conditions between 0.5 and 1.7 arcsec in the g band.  While cirrus emission has a deterrent impact on extragalactic studies -- its patchy structure is indeed difficult to subtract --, it  offers a powerful new tool to investigate the structure of the diffuse ISM and  constrain the process of energy dissipation.  With this database, we can probe a variety of cloud densities and   distances and  most likely  reach scales below 0.01\,pc. In the future, the Euclid mission will map a large part of the sky at  the unprecedented resolution of 0.1 arcsec, possibly offering an unexpected  new avenue for the study of the DGL.}

\begin{acknowledgements}
We thank the referee, Paul Goldsmith, for his careful reading of the manuscript.

The paper is based on observations obtained with MegaPrime/MegaCam, a joint project of CFHT and CEA/IRFU, at the Canada-France-Hawaii Telescope (CFHT) which is operated by the National Research Council (NRC) of Canada, the Institut National des Science de l'Univers of the Centre National de la Recherche Scientifique (CNRS) of France, and the University of Hawaii. 

The paper is also based on data from \Planck, a project of the European Space Agency (ESA) with instruments provided by two scientific consortia funded by ESA member states (in particular lead countries France and Italy), with contributions from NASA (USA) and telescope reflectors provided by a collaboration between ESA and a scientific consortium led and funded by Denmark.

The paper is also based on data from \Wise, the Wide-field Infrared Survey Explorer, a joint project of the University of California, Los Angeles, and the Jet Propulsion Laboratory/California Institute of Technology, funded by the National Aeronautics and Space Administration.
\end{acknowledgements}

\bibliography{draft_v5}
\bibliographystyle{aa}

\end{document}